\documentclass[runningheads]{llncs}
\usepackage[T1]{fontenc}
\usepackage{graphicx}
\begin{document}
\title{From Distorted Mirrors to Sovereign Reflections: Resisting the Grotesque Depiction of Our Digital Selves}
\titlerunning{From Distorted Mirrors to Sovereign Reflections}
%
\author{Andreas Komninos\inst{1}\orcidID{0000-0001-8331-1645} \and
Gerasimos Vonitsanos\inst{1}\orcidID{0000-0001-9555-4775} \and
Spyros Sioutas\inst{1}\orcidID{0000-0003-1825-5565}}
\authorrunning{A. Komninos et al.}
%
\institute{University of Patras, Rio 26504, Greece \\
\email{[akomninos, mvonitsanos, ssioutas]@ceid.upatras.gr}
}
\maketitle              
\begin{abstract}
As Ambient Intelligence weaves computing into everyday life, human existence has become inextricably linked to ubiquitous digital infrastructures, triggering a crisis of personal data sovereignty. Driven by extractivism and data colonialism, platforms exploit human experience as a terra nullius, enclosing proprietary user models within isolated silos. We argue that legal frameworks like the EU Digital Rights and GDPR fail to counter this power asymmetry because they regulate inert raw data while leaving monopolies over algorithmic inferences unchallenged. Consequently, current architectures act as funhouse mirrors, creating a grotesque, distorted depiction of our digital selves. Through critical posthumanism and four contemporary scenarios (entertainment, e-commerce, fintech, and telematics), this paper illustrates how data siloing inflicts cognitive amputation, ontological violence, and self-censorship. We argue that securing digital rights requires moving beyond defensive regulation toward a decolonial infrastructure that guarantees the functional transferability of algorithmic identities, reclaiming technology for the emancipation of the posthuman self.

\end{abstract}
\section{Introduction}

Technologists and futurists have anticipated Ambient Intelligence (AmI), a vision of a world where computing dissolves into the background of our lives \cite{cookAmbientIntelligenceTechnologies2009}. Today, this vision is no longer a speculative horizon, but rather constitutes our lived reality. We inhabit a landscape defined by ubiquitous data and service availability, where the environment itself is saturated with sensors, interconnected smart devices, and seamless digital infrastructures. From the smart homes that anticipate our morning routines to the connected vehicles that navigate our cities, technology has become an invisible, pervasive layer over the physical world, constantly harvesting data, predicting our behaviors, and seamlessly delivering services.

In this paradigm of ubiquitous computing, the boundary between the digital and the physical has entirely collapsed. We no longer "go online" to interact with technology, but rather, we are continuously bathed in it. Our biological existence and social functioning are now thoroughly intertwined with these ambient networks. However, this realization of Ambient Intelligence has triggered the profound and pressing crisis regarding digital rights and personal data sovereignty, foreseen in \cite{cookAmbientIntelligenceTechnologies2009}. Because computing is everywhere, the data extracted from us is also everywhere. Yet, the intelligence that animates our smart environments does not belong to us. As we navigate this ubiquitous landscape, fragmented and opaque algorithms construct proprietary \textit{"models-of-me"}: invisible, highly influential data doubles that dictate the services we receive, the prices we pay, and the opportunities we are afforded.

This creates a critical societal friction: we are completely dependent on a ubiquitous digital environment, but we possess zero functional agency over the algorithmic identities that represent us within it. When the environment's intelligence is locked in corporate silos, the lack of data sovereignty is not just a concern about online privacy, but an existential threat to our autonomy in the physical world. In this context, the assertion of digital rights, such as those outlined in the European Commission's digital principles \cite{DigitalRightsPrinciples}, constitutes an urgent, if insufficient in its own right, intervention to ensure that as technology becomes deeply interwoven with our human existence, we do not become subordinate to the ambient systems designed to serve us (and their creators).

In this paper, we present four scenarios that demonstrate the dissonance and power imbalance between citizens, as data producers and (theoretical) owners, and the organisations that leverage this data to provide the ubiquitous services on which our livelihoods depend. In exploring these scenarios, we argue that an alternative technological future of data sovereignty is not only possible, but highly desirable, if we are to escape the dangers brought forth by the current models of sociotechnical integration.

\section{Neocolonial Mythologies of Digital Rights and Data Sovereignty}
Modern societies have (or at least claim to have) made significant progress towards discarding the colonial legacy since the early enlightenment period. The colonial era justified the aggressive extraction of wealth from the Global South to the Global North, constructing deeply entrenched racial hierarchies. These societal architectures were contingent on the denial of primary human rights to those with least power, including the right of self-determination. Yet, although the era of  empires has largely ended, colonial architecture persists through modern neocolonial structures. The biases it created, remain embedded within global legal, social, and economic systems, perpetuating systemic discrimination and ensuring that wealthy nations and multinational corporations maintain disproportionate control over local material resources, at the expense of native populations \cite{mechkariniUnmaskingColonialMemory2023}. 

Where colonialism conquered through territory, the dominant socioeconomic model of neoliberalism dominates through the market. Neoliberalism demands the continuous expansion of the market into previously un-commodified areas of human life, economizing human existence and, in the process, eroding its vital rights \cite{astroulakisEthicalAnalysisNeoliberal2014}. Neoliberal capitalism's teleological drive toward the total commodification of existence  expands the logic of the market into spheres of human dignity, reducing intrinsic rights to tradeable values. As captured by Kant in 1785: \textit{"In the kingdom of ends everything has either a price or a dignity. What has a price can be replaced by something else as its equivalent; what on the other hand is raised above all price and therefore admits of no equivalent has a dignity"}. Protecting human rights thus requires both dismantling the lingering hierarchies of the colonial past and challenging the unfettered commodifying market dynamics of the present \cite{slaughterHijackingHumanRights2018,whyteHumanRightsCollateral2017}. 

Nowhere can one observe this dynamic more clearly, than in the market for digital services. In the digital age, the final frontier of privatization is human experience itself. This expansion relies on the mechanics of colonial computing \cite{breviniCritiquesDataColonialism2024,nothiasIntellectualHistoryDigital2025}. Historically, colonialism functioned through the violent appropriation of land, natural resources, and bodies from colonized populations, which were then processed by the imperial center into manufactured wealth and sold back to the dependencies. Data colonialism follows the exact same predatory logic \cite{doi:10.1177/1527476418796632,doi:10.1177/0263775816633195}. Big Tech views human life—our behaviors, emotions, health metrics, and social interactions—as a vast, unclaimed territory (terra nullius) ripe for extraction \cite{cohenBiopoliticalPublicDomain2019,magalhaesHumanLifeTerra2025}. The raw data is stolen or coerced from us, processed in the corporate "metropole" (the cloud servers of Silicon Valley or Beijing), and refined using proprietary AI into highly valuable predictive assets—the "models-of-me.". We are then forced to rent back the tools of our own social and biological reproduction (navigation, communication, healthcare management) from these digital empires \cite{doi:10.1177/08969205241302838}. We become structurally dependent on the very systems that exploit us. Worryingly, the model applies not just to private corporations, but increasingly so by governments, who see significant opportunities in the domination of digital spaces and the commodification (extraction and exploitation) of citizens' data, often in collaboration with private big-tech \cite{liuNavigatingDigitalColonialism2026}. 

In extant models of colonial computing, data silos created by the data each individual company or organisation collects on our human experience, are an economic necessity. They are the digital equivalent of colonial fences and borders. Just as early capitalism required the "enclosure of the commons" (fencing off public land to create private property), neoliberal computing requires the enclosure of data, allowing the harvesting and exploitation of what is termed the \textit{behavioral surplus} \cite{doi:10.1177/1095796018819461}. To maximize corporate valuation, companies must ensure that the algorithmic insights they extract from us, cannot escape their ecosystem. Therefore, technical infrastructure is deliberately engineered to prevent "functional transferability." The lack of interoperability is a feature of market domination, not a bug of software engineering.

This economic framing explains why legal compliance tools like the recently declared EU digital rights and GDPR are fundamentally inadequate, since while they provide ambitious intents, they lack enforceability and the ability for true citizen empowerment \cite{cocitoTransformativeNatureEU2023}. GDPR operates within a neoliberal paradigm of consumer rights and property ownership. It frames data as something we "own" and can choose to "trade" or "protect" via consent. Yet, data is not exactly property in the traditional sense, but instead, in digital capitalism, it is rather a "means of production" \cite{doi:10.1177/02673231241267128}. Further, in an asymmetric colonial relationship, "consent" is a myth. The average user cannot meaningfully withhold consent from the infrastructure required to live a modern life (e.g., banking, job applications, healthcare). Further, while GDPR refers to the right of data access and portability (Article 20), in practice, this usually means a user can download a static, incomprehensible CSV or JSON file of data like their raw search history, geolocation or music playback timelines. Theoretically, service providers can ingest this data into their platforms, allowing users to transfer from one provider to another, bringing their data with them (the "adieu" scenario \cite{dehertRightDataPortability2018}). Again, in practice, almost no provider caters for data migration due to technical boundaries such as lack of intelligibility via metadata uncertainties, cost of data transformation and missing data \cite{rubinfeldDataPortabilityInteroperability2024}.

Because neoliberalism privileges corporate power, global tech giants have the capital to absorb the costs of GDPR compliance. They employ armies of lawyers to turn compliance into a bureaucratic shield ("consent theater"), and while interoperability claims discuss the "de-siloisation" of data resources, in reality, new larger silos (platform silos) are created via data brokers, enabling large corporations to further entrench their power \cite{doi:10.1177/20539517241303144}. Meanwhile, the heavy compliance burden disproportionately crushes European SMEs and open-source alternatives \cite{hartingImpactsNewGeneral2021,sirurAreWeThere2018}. Paradoxically, GDPR has reinforced the monopolies of the data colonizers by locking out the competition. This is not, however, the largest problem. GDPR protects raw data, but the colonial value is extracted from algorithmic inferences. Because the law does not recognize that your "model-of-me" is an inseparable part of your posthuman self, it allows corporations to claim intellectual property rights over the AI models they trained on your life. Users cannot extract the utility of their data—the trained ML weights, the validated reputation, the nuanced preference profile, and know what the providers AI has inferred about them, but only the data points on which this inference was made. The law protects the colonizer's right to the refined product while giving the user rights only to the raw dirt. In other words, legal instruments like the GDPR leave users legally protected but functionally paralyzed \cite{turnerPrincipleVsPractice2024}. The law gives persons the right to hold their data, but not the power to use it, curate it, or direct its algorithmic destiny. 

\section{Mythology vs Reality}

Data silos assume that the raw data (e.g., a list of clicks, a heart rate log) is the valuable asset. In reality, raw data is useless without the algorithmic context that gives it meaning. The true value lies in the inferences drawn from that data, i.e. one's risk profile, aesthetic preferences, or health trajectory. Because these "models-of-me" are trapped in silos, our digital identity is constantly fragmented. As we move from our smart home to our connected car, or from our bank to a healthcare provider, our technological extensions cannot follow us. We are forced to constantly "re-build" our algorithmic selves from scratch in every new silo. This technical fragmentation prevents equitable access to society. If a verified, curated digital identity cannot be functionally transferred from one context to another, users are denied seamless access to cross-border or cross-domain digital public services. The user is always starting at a disadvantage, interacting with systems that do not truly "know" them, despite having extracted massive amounts of their data.

We argue that this not a failure of technology, as, after all, technical challenges have not been a major concern at least in the last decade. In the words of Bill Buxton, it's not a question of "if we can do" something, but rather, "how should we do it" \cite{uxlx-userexperiencelisbonUbiquitousComputingEmerging2014}. What leads to this dystopic, and frankly unnecessary situation, is the prevailing economic model of the digital market, built on the accumulation and monopolization of user data. Tech giants do not merely sell services, but trade in behavioral surplus and predictive models. In this economy, data silos are not accidental. Instead, they are a deliberate strategy to create insurmountable switching costs. If a user cannot take their highly tuned, personalized AI profile with them, they cannot afford to leave the incumbent platform. Because platforms hold a monopoly over a user's historical data and algorithmic profile, they no longer have to compete on the actual quality or ethical standing of their services. An emerging, privacy-respecting SME might offer an objectively better, fairer service, but it cannot compete if the user’s "model-of-me" is held hostage by a tech giant. The "starting user problem" is an artificial barrier that is often insurmountable and leads to failure. This economic model inherently violates liberal rights such as freedom of choice and self-determination of individuals. It transforms the digital economy into a technofeudal system where users are bound to the "estates" of a few global platforms \cite{durandTechnofeodalismeCritiqueLeconomie2020,varoufakisTechnofeudalism2024}. It stifles innovation and creates a market dynamic where the user's digital proxy is leveraged against them to maximize monopoly or oligopoly profits, stripping the user of agency and market power.

\section{The denied right to our digital self}
To fully understand why the friction surrounding personal data sovereignty is so profound, we must look beyond legal compliance (GDPR) and economic choices, and view it through the lens of critical posthumanism and technological interdependence. Posthumanism rejects the traditional humanist view that the human being is an isolated, autonomous entity entirely separate from nature and tools. Instead, philosophers like N. Katherine Hayles, Donna Haraway, and Bernard Stiegler argue that humans and technology co-constitute one another. We do not merely use technology; we evolve with it (technogenesis), and our cognition, identity, and social structures are inextricably woven into our technological systems \cite{ganeWhenWeHave2006,harawaySimiansCyborgsWomen2013,haylesHowWeBecame1999,TechnicsTime11998}. The situation of our proprietary "models-of-me" that lack "functional transferability," describes the corporate hijacking and alienation of the posthuman self. We now turn to some examples of this alienation, drawn from present-day lived experiences, to illustrate the situtation more concretely.

\subsection{Michael, the locked-in consumer}
Michael wants to switch from a dominant, global tech giant providing streaming music and audio service, to an emerging startup that promises fairer pay for creators and artists. Creating an account is easy enough. Yet, the new provider knows nothing about Michael's taste in music, since Michael cannot transfer across their curated playlists or collections metadata, or the analytics derived from the countless hours of use of the previous provider. Michael can legally use GDPR to download a .zip file of his raw listening history (a massive list of song titles in a CSV file). However, when he hands this raw file to the startup, it is useless because the startup doesn't have the proprietary algorithm to decode it, nor can Michael port the actual weights of his "taste profile." Not only does Michael need to rebuild their library from scratch, but the recommendation algorithm cannot be personalised until data from several hours of listening and interactions have been gathered on the new platform. Michael wonders if the 3€ per month in savings is worth the time he needs to invest in the new platform, to get the same level of service that he's used to with the current provider.

\subsection{Mary, the self-censored shopper}
Mary likes shopping online. She frequents a Europe-wide online marketplace to purchase items that are offered by private individuals and professional sellers, that often cannot be easily found in her home country. Yet a significant part of her purchases are made at a local online marketplace which mostly offers items from nationally located vendors. Mary prefers the latter when delivery speed is important, or when she wants to feel safer about making a potential return of purchased items, or the honouring of a guarantee. Mary's shopping behaviours are not shared between the platforms. As a result, product recommendations and sorting algorithms are only partially personalised in each platform. At the same time, Mary is aware of "polluting" the international platform's recommendation algorithm with data from search behaviour normally reserved for the other platform. She realises, through observation, that even a few searches for an item will result in days or weeks of being recommended related items to that search, and even though she is curious about being able to find bargains, she refrains from doing so to avoid being bombarded with recommendations that are likely irrevant outside the context of that fleeting curiosity. Even more, because the international platform monetises visits and searches through cross-site tracking, she knows that these irrelevant recommendations will appear as ads on many of the other websites she visits. Mary's problems increase when searches relate to potentially sensitive circumstances, for example looking for books on a recently diagnosed chronic illness, alternative political theories, or LGBTQ+ history. She becomes conscious of such searches, as this "fleeting curiosity" will permanently warp her algorithmic double across the entire web via cross-site tracking.

\subsection{Jimmy, the blinded entrepreneur}
Jimmy is wondering how a loan may help grow his small business. In the digital era, he doesn't need to visit local bank branches, but can shop for a loan online. The process is tedious - every time, he needs to enter all his personal details, upload financial statements and track records, proof of revenue, information on collateral assets, business formation documents and relevant business plans. He also needs to try out various options, such as payment scheduling, prepayment penalties, fixed costs and selecting between fixed and variable rates. Each bank has different requirements and he is happy to spend the time to provide this information, but there is one thing he cannot control: all bank require a credit score check, and Jimmy is new in business, so he doesn't have a great one. The "credit score" is an invisible, proprietary algorithmic tribunal. He cannot dispute it, he cannot audit it, and even more, he is unable to provide a projected score instead of his current one. Therefore Jimmy cannot use the banking websites to plan for when he might be eligible for a good deal on a loan, or the amount of money that he would be able to raise under sustainable terms. He is structurally blinded by the colonizer's code.

\subsection{Elena, the hyper-monitored commuter}
Elena participates in a "safe driver" program through a smartphone telematics app provided by her current auto insurance company, which monitors her phone's GPS and accelerometer to track her speed, hard braking, and phone distraction in exchange for a monthly discount. She also syncs her commercial fitness tracker with her health insurance provider's wellness app to earn premium rebates for meeting step goals. Recently, Elena’s auto insurance premium unexpectedly spiked. Because the app’s internal scoring model is completely opaque, she cannot see that the algorithm flagged her for "distracted driving" (when her passenger held her phone to navigate) and "high-risk behavior" (because she drives defensively and has had to break hard to avoid aggressive drivers or unexpected potholes in the road). Frustrated, Elena attempts to switch to a competitor. However, she discovers that her three years of data, which contains objective proof of her consistent sleeping habits, active lifestyle, and otherwise excellent driving metrics, is trapped in the proprietary databases of her current insurers. She cannot port the verified utility of her historical habits to the new provider. The competing insurance company treats her as a high-risk baseline, forcing her to choose between staying locked into an ecosystem that misinterprets and penalizes her behavior, or starting from scratch with a new corporate entity that will ambiently harvest her life all over again.

\section{Funhouse Mirrors of the Grotesque}
The 2022 European Declaration on Digital rights was published to guide the digital transition in line with EU values for a secure and sustainable digital future. The declaration centres on six themes: a) Putting people and their rights at the centre of the digital transformation; b) Supporting solidarity and inclusion; c) Ensuring freedom of choice online; d) Fostering participation in the digital public space; e) Increasing safety, security and empowerment of individuals (especially young people), and; f) Promoting the sustainability of the digital future. In part, these themes reflect a critical posthumanism perspective, in which, technology is not an external tool; it is an active co-constituent of human identity, cognition, and biology (technogenesis). When we map the European Commission’s digital rights principles onto the siloed, neoliberal, and colonial landscape in these scenarios, we see that the current paradigm does not just violate legal boundaries, but that it creates a profound ontological crisis through the grotesque transformation of our actual identities, when our digital self is viewed through these distorting, funhouse mirrors of fragmented digital doubles.

\subsection{Putting People at the Centre of the Digital Transformation}
Through the lens of critical posthumanism, the principle of putting people at the centre of the digital transformation demands that technology serves as an empowering extension of human agency (technogenesis), rather than treating the human as a mere biological host for data extraction. However, the scenarios illustrate a profound inversion of this principle under the current regime of data colonialism. In Elena’s case, her lived material reality (driving defensively to avoid hazards or having a passenger navigate) is entirely overwritten by the opaque, algorithmic double that her insurance provider constructs. The corporate technology does not serve her; rather, she is subordinated to the machine's misinterpretations, facing material financial punishment for organic behaviors that fail to align with the platform's predictive models. Jimmy is reduced to an inauditable credit score, a proprietary asset of the bank rather than a reflection of his actual entrepreneurial potential. In these neoliberal architectures, the human is not the center of the transformation, but merely a raw resource to be strip-mined for behavioral surplus. Consequently, individuals are forced to adapt their physical behaviors to appease the algorithmic caricatures that govern their lives, entirely stripping them of central agency.

\subsection{Solidarity and Inclusion}
In a deeply interconnected posthuman society, solidarity and inclusion require that the technological infrastructures mediating our lives facilitate equitable access and bridge structural divides. The current paradigm of data siloing weaponizes our digital interdependence to automate and deepen exclusion. Jimmy’s inability to access sustainable financial capital stems from a systemic refusal to recognize his holistic financial history across different platforms; because he cannot functionally transfer his verified data to prove his viability, the algorithmic tribunal defaults to exclusionary prejudice based on his status as a new business owner. Elena faces a similar fracture. Her historically validated safe behaviors and active lifestyle are rendered invisible simply because she crossed a corporate border by seeking a new provider. By treating her as a high-risk baseline, the new ecosystem effectively penalizes her lack of algorithmic capital within its specific walled garden. Under surveillance capitalism, digital services are driven by predictive extractivism rather than human well-being, meaning that those without perfectly portable, platform-specific algorithmic histories, are structurally pushed to the margins, shattering social solidarity.

\subsection{Freedom of Choice}
The principle of freedom of choice is conceptually void when analyzed through the posthumanist framework of the Extended Mind Thesis, which posits that our tools and digital environments are literal, externalized extensions of our cognition and memory. Michael’s scenario captures how neoliberal data architectures engineer artificial lock-in to eliminate true market competitiveness. Michael’s curated playlists, listening history, and resulting algorithmic taste profile are not merely inert data points; they represent the externalization of his aesthetic identity. When the dominant tech giant prevents the functional transferability of his "model-of-me," Michael's legal right to download a raw data file under the GDPR becomes functionally meaningless. To switch providers, he must suffer a form of cognitive amputation, abandoning a deeply integrated part of his digital consciousness. Thus, his freedom of choice is revealed to be a coercive illusion. He is economically and psychologically bound to the incumbent data colonizer because the infrastructural architecture makes the cost of reclaiming his digital self—rebuilding his identity from scratch—prohibitively high.

\subsection{Participation in the Digital Public Space}
Meaningful participation in the digital public space necessitates an environment where individuals can freely explore ideas, access information, and engage in the fluid process of posthuman, Hegelian \textit{becoming} without fear of algorithmic reprisal. Mary’s scenario exposes how the privatization and panoptic surveillance of digital spaces actively freeze this identity evolution. Because her online environment is driven by cross-site tracking and the relentless monetization of attention, Mary is acutely aware that any deviation from her established routine, such as researching a chronic illness or exploring LGBTQ+ history, will permanently warp her algorithmic double. This induces a profound chilling effect. Mary is forced into self-censorship, actively policing her own curiosity to avoid "polluting" her recommendation algorithms and being hunted by irrelevant or sensitive advertisements across the web. When the digital public space operates as a behavioral market, participation becomes conditional upon conforming to predictable patterns, stripping the posthuman subject of the freedom to safely simulate alternative interests and effectively locking them in a prison of their past behaviors. In Hegel's terms, Mary is limited to a static state of \textit{being}. Denied her potential for \textit{becoming}, both she, and the platforms, negate the ability to grasp the foundational structure of her reality (\textit{Geist}).

\subsection{Safety, Security, and Empowerment}
Current legal frameworks often reduce safety and empowerment to defensive consumer rights, such as ticking cookie consent banners or relying on nominal data protection clauses. Critical posthumanism, however, reveals that true security requires absolute functional sovereignty over the cyborg assemblage—the hybrid of flesh, data, and code that constitutes the modern subject. Jimmy and Elena are subjected to ontological violence precisely because they lack this empowerment. Their material survival and financial security depend entirely on digital representations over which they possess zero functional agency. Jimmy cannot audit the algorithmic judgment of his creditworthiness, nor can he project future scenarios to guide his business decisions. He is structurally blinded by the colonizer's code. Elena cannot dispute the false positives generated by her telematics app. Without user-managed environments that act as a sovereign "Personal Mirror", allowing individuals to audit, curate, and dynamically negotiate how institutional algorithms perceive them, humans remain structurally defenseless against the pervasive, automated judgments of ambient intelligence.

\subsection{Sustainability}
While often framed purely in ecological terms, the EU principle of sustainability is fundamentally incompatible with the digital architectures of data colonialism when viewed through a posthuman lens. The deliberate obstruction of functional transferability forces a continuous, redundant extraction and processing of the human experience. Because Michael cannot port his taste profile, the new streaming startup must expend immense computational energy to re-harvest his behavioral surplus and train a new predictive model from scratch. Likewise, because Elena’s verifiable driving history is trapped in a proprietary database, her new insurance provider must "ambiently harvest her life all over again," deploying redundant sensor tracking and algorithmic processing to establish a profile that already exists elsewhere. This forced duplication of algorithmic labor, engineered solely to maintain artificial corporate monopolies, results in a grotesque squandering of computational power, data storage, and electrical energy. The neoliberal enclosure of the digital self therefore directly links the exploitation of human identity to the unsustainable exhaustion of the planetary ecology.

\begin{figure}
    \centering
    \includegraphics[width=1\linewidth]{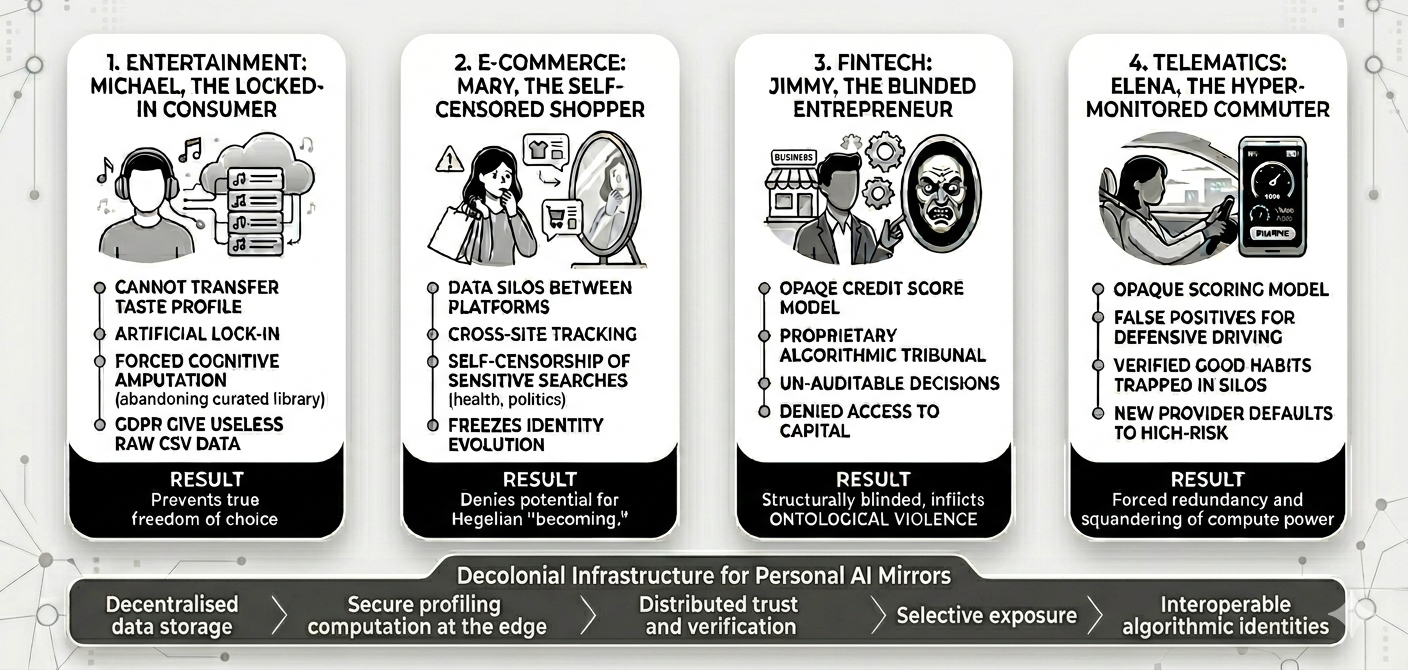}
    \caption{The current issues with data siloing and a proposed alternative technological future. Infographic composed with Gemini 3.1 Pro.}
    \label{fig:figure}
\end{figure}

\section{Another Technological Future}

The current architecture of surveillance capitalism often presents itself as the only technologically viable method for delivering ambient intelligence. However, the centralized, extractivist model of data siloing is an economic design choice rather than an infrastructural inevitability. An alternative paradigm is not only technologically feasible but urgently desirable if we are to dismantle the neocolonial structures of the digital economy and restore posthuman agency. This alternative requires a fundamental infrastructural inversion: shifting the locus of control from corporate cloud metropoles back to the individual through a combination of edge-computing architectures, decentralized ledger technologies, and cryptographic privacy frameworks (Figure \ref{fig:figure}).

The technological foundation of this alternative lies in the distributed storage of data across a user’s own network of personal devices, utilizing decentralized storage networks or secure Trusted Execution Environments (TEEs) embedded directly within consumer hardware. In this decentralized paradigm, the vast streams of ambient data generated by a user's daily life—spanning from fitness wearables and smart home sensors to localized media consumption and financial ledgers—are retained within a secure, user-owned perimeter. Rather than exporting raw behavioral surplus to centralized corporate servers to be synthesized into proprietary "models-of-me," the computation occurs locally at the edge. By utilizing decentralized identity (DID) standards and blockchain architectures, the system can establish a tamper-proof registry of data provenance and consent without storing any raw personal information on a public ledger. Blockchain functions here purely as a decentralized infrastructure for trust, logging immutable cryptographic hashes that verify the integrity of the local data sources and the compliance of interacting third parties.

Crucially, this architecture operationalizes the principle of functional transferability by shifting the point of exchange from raw data extraction to new protocols for secure information exchange and cryptographic disclosure. Rather than demanding granular, raw data to train corporate predictive models, service providers interact with a universal interoperability layer—conceptualized as the "TCP/IP of Personal AI." This layer utilizes Zero-Knowledge Proofs (ZKPs), such as zk-SNARKs, to securely expose only the validated semantic outcomes of localized computation. Under this framework, the user's personal system computes the profile once, and subsequently generates a mathematical proof verifying the semantic result without revealing the underlying raw data. For example, Elena's system can cryptographically prove she meets a specific "safe driver" threshold without disclosing her exact GPS coordinates, while Jimmy's system can verify a creditworthiness benchmark without transferring years of raw corporate transactions. Through these privacy-preserving protocols, users selectively expose context-specific inferences to service providers, leaving businesses unable to capture or hoard the underlying human experience.

To ensure that this cryptographic complexity does not alienate the human subject, the architecture requires a sophisticated, user-facing semantic interface, a sovereign "Personal Mirror." This mirror serves as a trustworthy, independently verifiable dashboard that translates raw machine learning metrics and ZKP outputs into a legible representation of the user's digital identity. For the mirror to be truly trustworthy, it must integrate uncertainty quantification metrics, enabling the system to communicate its own algorithmic confidence levels transparently to the user, and adversarial robustness testing to ensure the profile has not been poisoned by external corporate manipulation. The independent verifiability of this mirror is guaranteed through local cryptographic signatures; the user can verify that the mirror’s depiction is an untampered, mathematically accurate reflection of their local data state, completely decoupled from tracking.

Through this interactive dashboard, humans are empowered to actively curate their data and select specific items to compose context-specific algorithmic profiles, or "AI Outfits." A user can review the various facets of their externalized digital self, dynamically deciding which verified historical attributes or semantic inferences to package together for a specific interaction. Michael can seamlessly bundle his verified taste profiles and playlist metadata into a portable musical identity to hand to a new startup, while Mary can instantiate a temporary, anonymized browsing profile that shields her fluid intellectual curiosity from cross-site ad networks. By allowing individuals to dynamically assemble, audit, and authorize these algorithmic outfits, the Personal Mirror transforms the posthuman subject from a passive object of corporate extractivism into an active curator of their own digital destiny.

The desirability of this model extends beyond individual privacy; it promises a radical and necessary restructuring of the digital economy by shifting the competitive landscape away from algorithmic dominance and back toward genuine service provision. In the current neoliberal paradigm, corporate monopolies are sustained by artificially locking users into walled gardens and fiercely guarding massive silos of behavioral data, meaning that the scale of data ownership dictates market supremacy. However, if users bring their own pre-computed, universally interoperable profiles to the market via secure disclosure protocols, this artificial lock-in evaporates. Businesses would no longer be able to compete based on who possesses the largest data silo or the most exclusively trained predictive model. Instead, the algorithmic playing field is leveled, forcing platforms to compete purely on the material value, quality, pricing, and ethical standing of the actual services they deliver. Ultimately, this decentralized approach dismantles the grotesque, distorted mirrors of data colonialism, ensuring that technology ceases to function as a tool of ambient subjugation and reclaims its place as an empowering, manageable, and liberating extension of human agency.

%
%
%
\bibliographystyle{splncs04}
\bibliography{refs}

\end{document}